\documentclass[journal=amlcef,manuscript=letter]{achemso}

\usepackage{graphicx}
\usepackage{dcolumn}
\usepackage{bm}
\usepackage[version=4]{mhchem}
\usepackage{gensymb}
\usepackage{chemformula} 
\usepackage[T1]{fontenc} 
\usepackage{textgreek} 
\usepackage{amsmath}
\usepackage{amssymb}
\usepackage{color,soul}
\usepackage{textcomp}
\usepackage{bm}
\usepackage[normalem]{ulem}
\usepackage[colorlinks=true,linkcolor=blue,urlcolor=blue]{hyperref}

\title{\ce{GdWN3} is a Nitride Perovskite}

\author{Rebecca W. Smaha}
\email{Rebecca.Smaha@NREL.gov}
 \affiliation{National Renewable Energy Laboratory, Golden, Colorado 80401, USA} 
\author{John S. Mangum}
 \affiliation{National Renewable Energy Laboratory, Golden, Colorado 80401, USA} 
   \author{Neha Yadav}
\affiliation{Colorado School of Mines, Golden, Colorado 80401, USA} 
\author{Christopher L. Rom}
 \affiliation{National Renewable Energy Laboratory, Golden, Colorado 80401, USA} 
 \author{Brian M. Wieliczka}
 \affiliation{National Renewable Energy Laboratory, Golden, Colorado 80401, USA} 
 \author{Baptiste Julien}
 \affiliation{National Renewable Energy Laboratory, Golden, Colorado 80401, USA} 
  \author{Andrew Treglia}
 \affiliation{Colorado State University, Fort Collins, Colorado 80521, USA} 
 \author{Craig L. Perkins}
 \affiliation{National Renewable Energy Laboratory, Golden, Colorado 80401, USA} 
 \author{Prashun Gorai}
\affiliation{Colorado School of Mines, Golden, Colorado 80401, USA} 
 \alsoaffiliation{National Renewable Energy Laboratory, Golden, Colorado 80401, USA} 
 \alsoaffiliation{Rensselaer Polytechnic Institute, Troy, New York 12180, USA}
 \author{Sage R. Bauers}
 \affiliation{National Renewable Energy Laboratory, Golden, Colorado 80401, USA} 
\author{Andriy Zakutayev}
\email{Andriy.Zakutayev@NREL.gov}
 \affiliation{National Renewable Energy Laboratory, Golden, Colorado 80401, USA} 

\begin{document}

\begin{abstract}
Nitride perovskites \ce{\textit{AB}N3} are an emerging and highly under-explored class of materials that are of interest due to their intriguing calculated ferroelectric, optoelectronic, and other functional properties. Incorporating novel $A$-site cations is one strategy to tune and expand such properties; for example, \ce{Gd^{3+}} is compelling due to its large magnetic moment, potentially leading to multiferroic behavior. However, the theoretically predicted ground state of \ce{GdWN3} is a non-perovskite monoclinic structure. Here, we experimentally show that \ce{GdWN3} crystallizes in a perovskite structure. High-throughput combinatorial sputtering with activated nitrogen is employed to synthesize thin films of \ce{Gd_{1-x}W_{x}N_{3-y}} with low oxygen content within the bulk of the films. Ex-situ annealing crystallizes a polycrystalline perovskite phase in a narrow composition window near $x=1$. LeBail fits of synchrotron grazing incidence wide angle X-ray scattering data are consistent with a perovskite ground-state structure. New density functional theory calculations that included antiferromagnetic configurations confirm that the ground-state structure of \ce{GdWN3} is a distorted $Pnma$ perovskite with antiferromagnetic ordering, in contrast to prior predictions. Initial property measurements find that \ce{GdWN3} is paramagnetic down to $T=2$ K with antiferromagnetic correlations and that the absorption onset depends on cation stoichiometry. This work provides an important stepping stone towards the rapid expansion of the emerging family of nitride perovskites and towards our understanding of their potential multiferroic properties.
\end{abstract}

\maketitle

\newpage

Nitride perovskites \ce{\textit{AB}N3} have been of interest in recent years due to the highly underexplored nature of this phase space and the exciting piezoelectricity observed for \ce{LaWN3}.\cite{Talley2021}  To date, only five nitride perovskites have been reported experimentally: \ce{TaThN3},\cite{Brese1995} \ce{LaWN3},\cite{Talley2021,MATSUISHI2022} \ce{LaReN3},\cite{Klos2021} \ce{CeWN3}, and \ce{CeMoN3}.\cite{Sherbondy2022} However, theoretical calculations have predicted that a wide phase space of nitride perovskites should exist with highly oxidized \textit{B} cations such as \ce{W^{6+}}, \ce{Re^{6+/7+}}, \ce{Mo^{6+}}, \ce{Ta^{5+}}, \ce{Nb^{5+}}, \ce{Cr^{6+}}, or \ce{Os^{6+/8+}}.\cite{Sarmiento-Perez2015,Korbel2016,Flores-Livas2019,Ha2022,Sherbondy2022,Grosso2023} While they are predicted to be stable, in practice it is challenging to synthesize these materials and ensure that they are fully nitrided. Many oxynitride perovskites (e.g., \ce{\textit{AB}O_{1+x}N_{2-x}} \cite{Fuertes2012}) and nitrogen-poor antiperovskite nitrides (i.e., \ce{\textit{M}3N\textit{E}} \cite{Niewa2019}) have been reported, demonstrating that the challenge is greater for nitride-rich materials. The most successful techniques towards stoichiometric nitride perovskites have been radio-frequency sputtering,\cite{Talley2021,Sherbondy2022} which increases the chemical potential of nitrogen and is a highly energetic process, and high-pressure high-temperature methods.\cite{Klos2021,MATSUISHI2022} In addition to phase stability, calculations predicted that several rare earth nitride perovskites should exhibit ferroelectric behavior and be semiconducting.\cite{Sarmiento-Perez2015} \ce{LaWN3} was synthesized via reactive sputtering and shown to have a large piezoelectric response and thus polar symmetry.\cite{Talley2021} However, its physical properties are highly sensitive to stoichiometry, necessitating precise synthetic control in order to functionalize this material---as is the case for many semiconductors.\cite{Smaha2023} 

Expanding the palette of $A$ cations can expand the potential functionality of nitride perovskites. One attractive target is multiferroicity, in which a material exhibits ferroelectricity and magnetic order simultaneously. They are an emerging option for electronic applications such as low-power field sensing, memory recording/storage technologies, and spintronic devices.\cite{fiebig2016,Spaldin2019} However, achieving both ferroic orders simultaneously is difficult: ferroelectricity normally arises from an ionic displacement that breaks inversion symmetry, but this is generally related to a d$^0$ electronic structure or a lone pair, precluding magnetic order in transition metals such as Fe, Mn, etc.\cite{Spaldin2019} To this end, $A$ = \ce{Gd^{3+}} is an attractive option, as \ce{Gd^{3+}} has one of the largest magnetic moments (8 $\mu_B$), and elemental Gd is unique among the rare earths in that it is ferromagnetic (FM) at room temperature. Recent calculations examined the ground-state structure of \ce{GdWN3} but did not consider the possible antiferromagnetic (AFM) ordering of \ce{Gd^{3+}}, which is crucial for the correct identification of the ground-state structure. Unfortunately, Flores-Livas et al. and Gui et al. calculated a non-perovskite monoclinic (\textit{C2/c}) ground-state structure.\cite{Flores-Livas2019,Gui2022}  However, additional magnetic structure calculations of two perovskite phases that were slightly higher in energy than the monoclinic ground state indicated large ferroelectric polarization and AFM order (albeit with N\'eel temperature $T_N<8$ K), raising the possibility of stabilizing a single-phase multiferroic material.\cite{Gui2022}

Here, we conclusively show that \ce{GdWN3} can be experimentally realized as a nitride perovskite using a combination of high-throughput synthesis and characterization. We employ combinatorial sputtering with activated nitrogen to survey the \ce{Gd_{1-x}W_{x}N_{3-y}} phase space and use a variety of electron microscopy and X-ray techniques to probe its composition and crystal structure. Strikingly, a perovskite phase crystallizes close to the \ce{GdWN3} composition. Confirming the experimental results, we use first-principles calculations to compute that a perovskite phase (\textit{Pnma}) with antiferromagnetic ordering is the ground-state structure of \ce{GdWN3}. We perform initial property measurements, finding that \ce{GdWN3} is paramagnetic down to $T=2$ K, in contrast to previous computational predictions, but exhibits AFM correlations. In addition, there is a large change in optical properties across the \ce{Gd_{1-x}W_{x}N_{3-y}} composition gradient. This work yields new understanding of the thermodynamics within---and significantly expands---the small phase space of experimentally reported nitride perovskites.

Combinatorial deposition of \ce{GdWN3} was performed with RF co-sputtering onto substrates held at high temperature ($\sim$700 $^\circ$C), yielding compositionally-graded films. Throughout this article, we use \ce{GdWN_{3-y}} to refer to the combinatorial films and specify the cation composition Gd/(W+Gd) at measured points; other nitride perovskites grown as films generally have some amount of N deficiency ($y\approx0.5$), as noted by Ref. \cite{Talley2021}. Cation composition Gd/(W+Gd) was mapped across the combinatorial samples using X-ray fluorescence (XRF), and several points were additionally measured with Auger electron spectroscopy (AES) sputter depth profiling and scanning transmission electron microscopy (STEM) energy-dispersive X-ray spectroscopy (EDS) to probe the anion stoichiometry, as discussed further below. The growths yielded films with a composition gradient roughly $30\% < $ Gd/(W+Gd) $< 60\%$ (i.e., \ce{Gd_{0.6}W_{1.4}N_{3-y}} to \ce{Gd_{1.2}W_{0.8}N_{3-y}}). 

\begin{figure}
\includegraphics[width=1\textwidth]{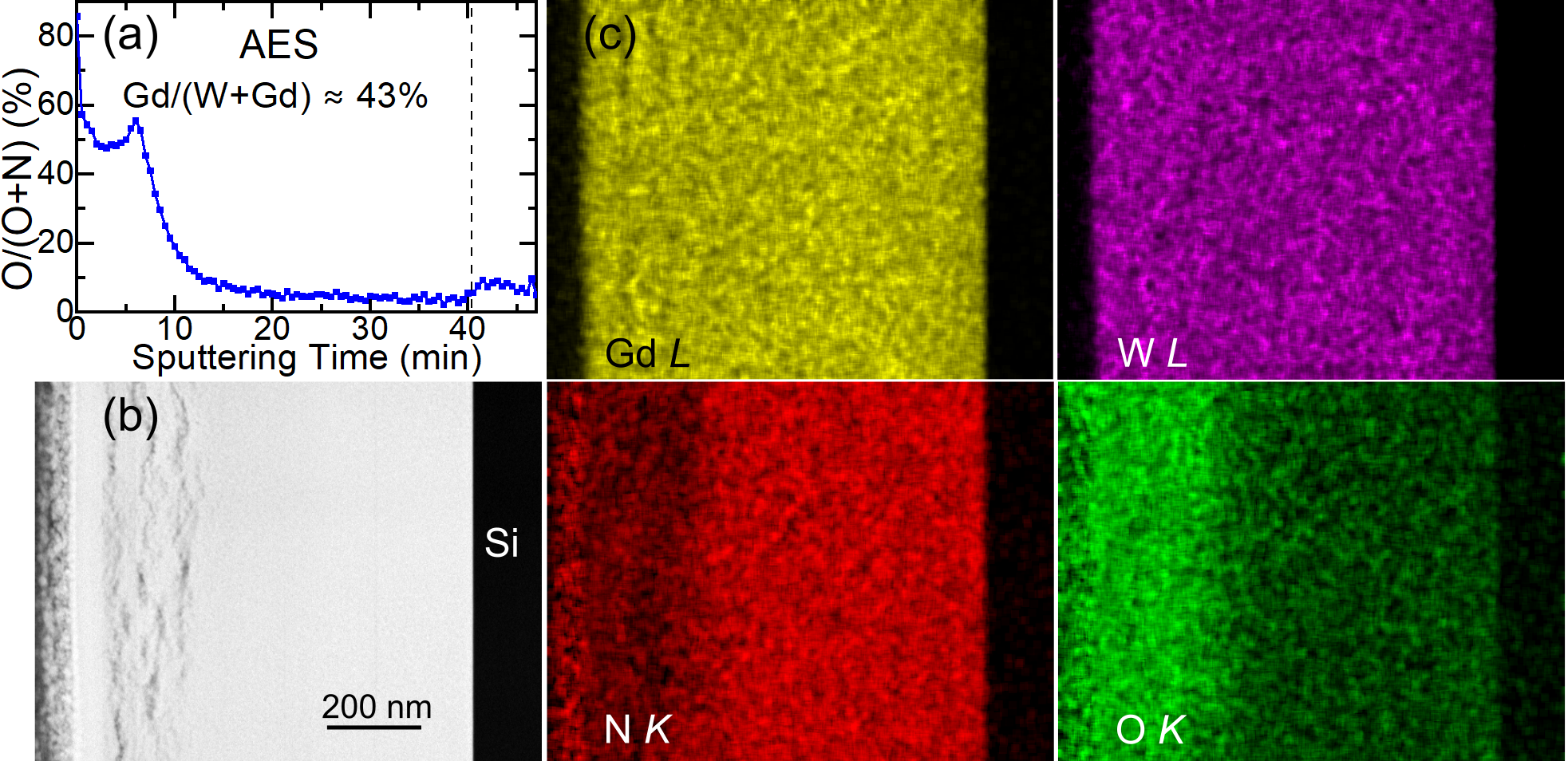}\\
  \caption{(a) Cation ratio O/(N+O) extracted from the AES sputter depth profile of this sample. (b) STEM HAADF image aligned with the AES depth profile. (c) STEM-EDS elemental mapping at the Gd \textit{L}-peak, W \textit{L}-peak, N \textit{}K-peak, and O \textit{K}-peak. All measurements were performed on the same region of a combinatorial film with Gd/(W+Gd) $\approx$ 43\%.} 
  \label{fgr:TEM}
\end{figure}

AES sputter depth profiling performed on a sample region with Gd/(W+Gd) $\approx$ 43\% (Fig. \ref{fgr:TEM}a) shows an oxygenated surface layer, potentially implying the presence of some amorphous Gd oxide (\ce{GdO_x}). However, it also indicates low O content in most of the film, closer to $\sim$5\% O/(N+O). 4D-STEM results (Fig. S2) show that the grain size ranges from approximately 130 -- 200 nm. 
STEM high-angle annular dark-field (HAADF) imaging and EDS elemental mapping were performed on this sample (Fig. \ref{fgr:TEM}b,c). The STEM-EDS mapping exhibits no obvious segregation of Gd and W throughout the film, but the analyzed film shows clear signs of oxidation due to air exposure post-deposition: additional O is visible near the surface. Near the surface, the film has a large amorphous region that is oxygen-rich, consistent with the AES and STEM-EDS results and likely due to post-deposition oxidation (Fig. \ref{fgr:TEM}a and Fig. S2). In addition, a few laterally-aligned voids that may be delamination induced by an ex-situ rapid thermal annealing step are visible in Fig. \ref{fgr:TEM}a. A scanning electron microscopy (SEM) image of a spot along a combinatorial film grown on a Si substrate after annealing is shown in Fig. S3 and exhibits similar laterally-aligned voids near the top surface of the film. AES profiling performed on a freshly deposited sample (Fig. S1) exhibits similarly low O content in the bulk (2 -- 4\%) and a thinner region of higher O near the surface, confirming that the O seen in the sample analyzed with STEM is from post-growth oxidation, potentially including from the ex-situ annealing step and/or oxidation of the lamella. 

\begin{figure}[h]
\includegraphics[width=0.5\textwidth]{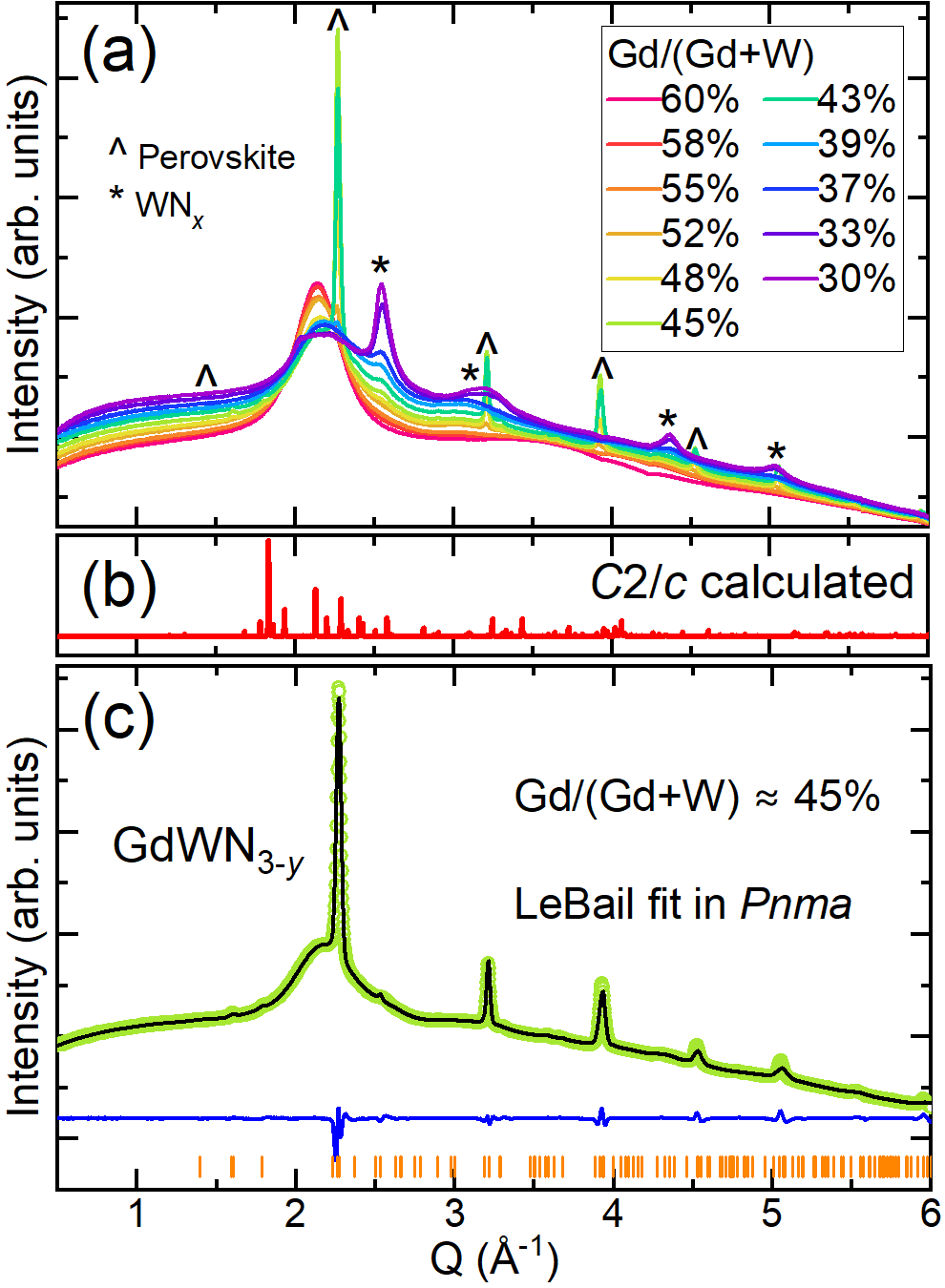}\\
  \caption{(a) Synchrotron grazing incidence wide angle X-ray scattering (GIWAXS) data along a compositionally-graded \ce{GdWN_{3-y}} film. (b) Simulated pattern of the previously calculated monoclinic \textit{C2/c} ground-state structure. (c) LeBail fit of the synchrotron GIWAXS dataset with Gd/(W+Gd) $\approx$ 45\% in the calculated ground state structure perovskite space group $Pnma$. The data (green circles), fit (black line), and residual (blue line) are shown. The cation ratio Gd/(W+Gd) was measured with XRF.} 
  \label{fgr:XRD}
\end{figure}

Ex-situ rapid thermal annealing in flowing \ce{N2} causes the as-deposited amorphous or nanocrystalline films to crystallize near the \ce{GdWN3} composition. Laboratory X-ray diffraction (XRD) data of an as-deposited film and the same film annealed at 950 $^\circ$C in flowing \ce{N2} are shown in Fig. S4.  Synchrotron grazing incidence wide angle X-ray scattering (GIWAXS) was employed to study the structure of this annealed film along the composition gradient, as shown in Fig. \ref{fgr:XRD}a. The combinatorial GIWAXS data show that the width of the new phase in composition space is narrow, less than $10$\% Gd/(W+Gd), and the highest crystallinity is not found at Gd/(W+Gd) $=50$\% but rather at a slightly Gd-poor value of Gd/(W+Gd) $\approx 45$\%. In addition to this new phase, a broad amorphous peak at $Q\approx2.1$ \AA$^{-1}$ is often observed on the Gd-rich side of the films after annealing, likely arising from a \ce{GdO_x} impurity. On the Gd-poor side of the films, a broad peak arising from \ce{WN_x} is visible at $Q \approx 2.6$ \AA$^{-1}$. In some films, metallic W and/or \ce{GdO_x} crystallized on the W- and Gd-rich sides of the film (Figs. S4 and S7). 

Comparing the peaks of the unidentified crystalline phase to the simulated pattern of the previously calculated monoclinic \textit{C2/c} ground-state structure (Fig. \ref{fgr:XRD}b) does not yield a match. Instead, the pattern is well-matched by a perovskite phase, as shown by the carets in Fig. \ref{fgr:XRD}a. GIWAXS 2D detector images (Fig. S5) indicate that this perovskite phase is polycrystalline. To confirm the unexpected discovery of a perovskite structure, we performed LeBail fits on GIWAXS data of a film region with Gd/(W+Gd) $\approx$ 45\% in three possible perovskite space groups ($Pm\bar{3}m$, $R3c$, and $Pnma$). Orthorhombic $Pmna$ yields the best fit, shown in Fig. \ref{fgr:XRD}b; fits to polar $R3c$ (the space group in which \ce{LaWN_{3-y}} crystallizes\cite{Talley2021,MATSUISHI2022,Smaha2023}) and the prototypical cubic perovskite space group $Pm\bar{3}m$ are statistically equivalent to each other and shown in Fig. S6. A summary of the extracted lattice parameters and fit metrics is in Table S1. We note that the annealing temperature required to crystallize the \ce{GdWN_{3-y}} perovskite phase is 900 -- 950 $^\circ$C, slightly higher than that required to crystallize \ce{LaWN_{3-y}} ($\sim$800 $^\circ$C),\cite{Talley2021,Smaha2023} and both the narrowness of the phase width and the most crystalline composition being \textit{A}-poor are consistent with previous results seen for \ce{LaWN_{3-y}}.\cite{Smaha2023} 

\begin{figure}[h]
\includegraphics[width=\textwidth]{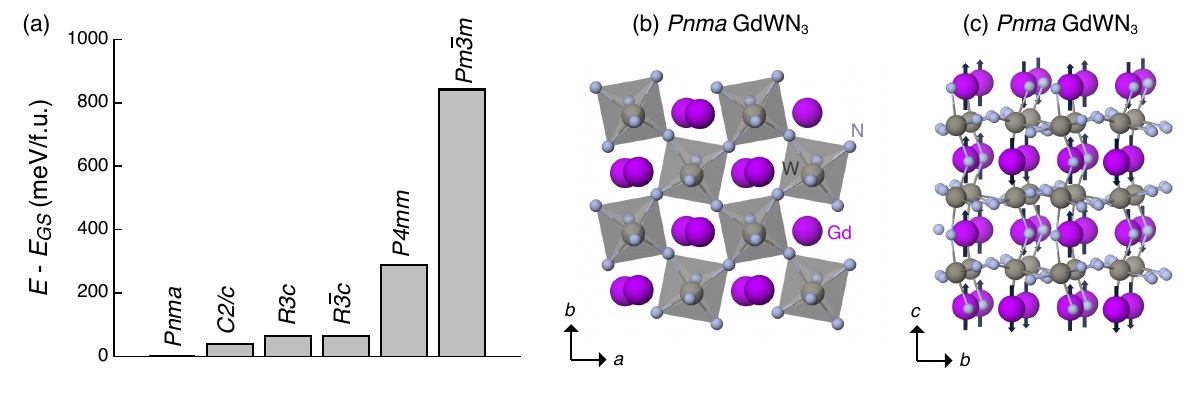}\\
  \caption{(a) Energies of \ce{GdWN3} polymorphic structures relative to the predicted ground-state \textit{Pnma} structure. \textit{C2/c} is not a perovskite phase, while the other structures are in the perovskite family. (b) Stable ground-state \textit{Pnma} structure. (c) Antiferromagnetic ground-state magnetic structure of \textit{Pnma} \ce{GdWN3}.}
  \label{fgr:calc}
\end{figure}

As our experimental results are inconsistent with previous computational predictions of the structure of \ce{GdWN3}, we performed new density functional theory (DFT) calculations to search for the magnetic ground-state structure of \ce{GdWN3}. We considered six polymorphic structures (both perovskite and non-perovskite phases) in the following space groups: perovskites \textit{Pm$\bar{3}$m}, \textit{Pnma}, \textit{P4mm}, \textit{R3c}, and \textit{R$\bar{3}$c}, and non-perovskite \textit{C2/c} (the  structure previously calculated as the ground state\cite{Flores-Livas2019,Gui2022}); see Supporting Information for details. Considering only FM ordering, the ground-state energy of the perovskite \textit{Pnma} and the non-perovskite \textit{C2/c} phases are similar (Table S2), which agrees with prior computational work that found monoclinic \textit{C2/c} to be the ground-state structure.\cite{Flores-Livas2019,Gui2022} Next, we considered various AFM configurations of \ce{Gd^{3+}} to search for magnetic orderings that are lower energy than FM, as the huge magnetic moment of Gd$^{3+}$ must play an important role in phase stability. Interestingly, we find a distorted perovskite phase in orthorhombic \textit{Pnma} (Fig. \ref{fgr:calc}b,c) to be a lower-energy structure for \ce{GdWN3}. The relative energies of the polymorphic structures in their lowest-energy magnetic ordering are shown in Fig. \ref{fgr:calc}a and Table S2, and the relaxed structures are shown in Fig. S9. Cubic perovskite \textit{Pm$\bar{3}$m} has the highest energy in comparison to the other phases, implying that symmetry-lowering, including octahedral distortions and tilting, stabilizes the other perovskite phases (\textit{Pnma}, \textit{P4mm}, \textit{R3c}, \textit{R$\bar{3}$c}). The calculated and experimental lattice parameters extracted from LeBail fits of three of the possible space groups are in close agreement, with the best agreement for the \textit{Pnma} phase, further indicating that this phase is the ground-state structure (Table S1). While these data and refinements confirm that \ce{GdWN_{3-y}} crystallizes within the perovskite family, future work will be needed to definitively assign the symmetry. 

\begin{figure}
\includegraphics[width=0.5\textwidth]{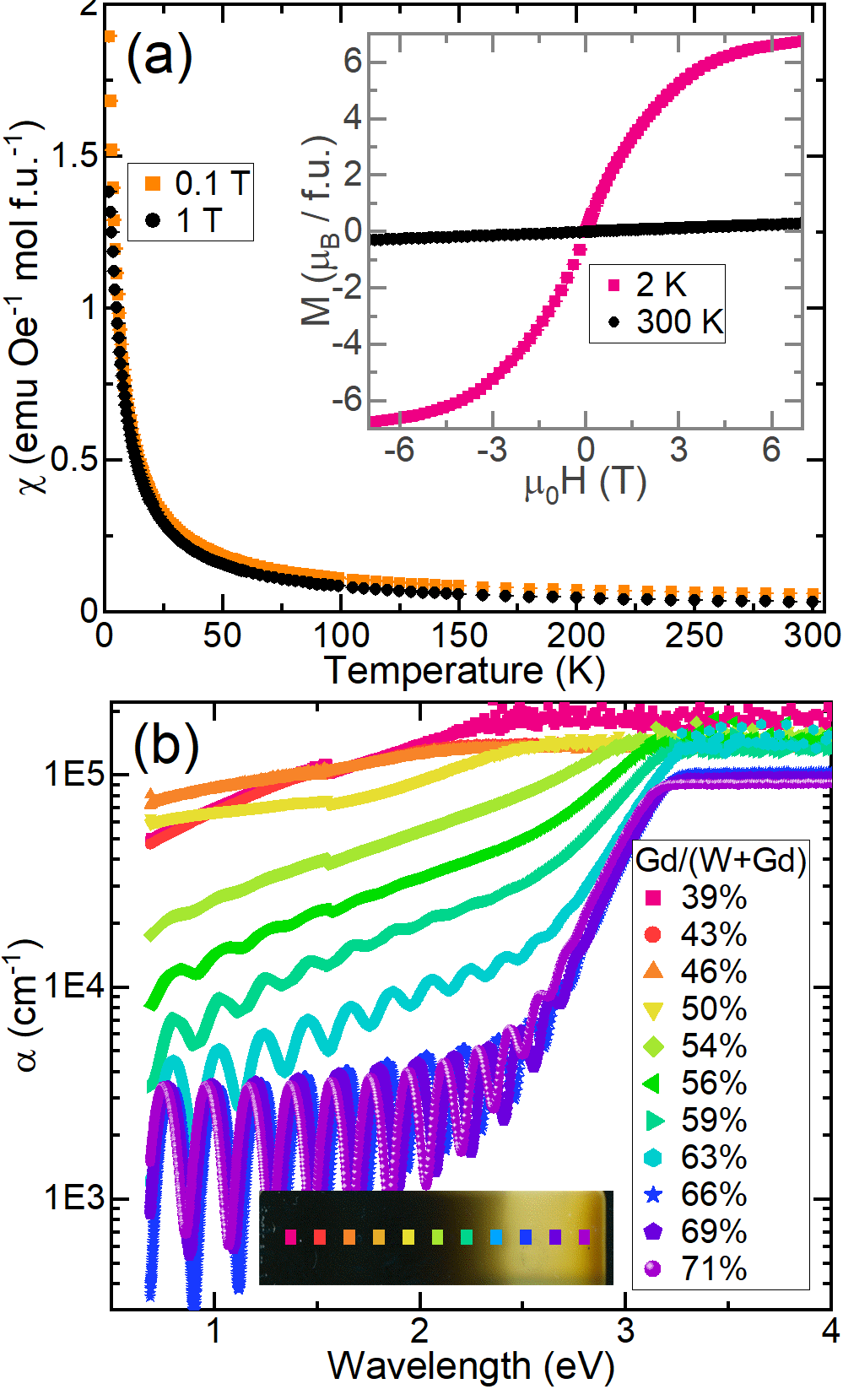}\\
  \caption{(a) Magnetic susceptibility ($\chi$) of \ce{GdWN_{3-y}} as a function of temperature at several different applied fields. The inset shows magnetization ($M$) as a function of applied field. The measured film region had Gd/(W+Gd) $\approx$ 43\%. (b) Absorption coefficient ($\alpha$) extracted from UV-vis measurements along a compositionally-graded \ce{GdWN_{3-y}} film. The inset shows a picture of this film, and the approximate measured locations are marked.} 
  \label{fgr:prop}
\end{figure}

Our DFT calculations predict an AFM ground state for \textit{Pnma} \ce{GdWN3} with the Gd$^{3+}$ moments arranged in planes along the $c-a$ axis and anti-parallel in adjacent planes stacked along the $b$ axis (Fig. \ref{fgr:calc}c). This is similar to previous calculations of an AFM ground state with $T_N>2$ K for \textit{R3c} and \textit{Pna2$_1$} perovskite phases.\cite{Gui2022} To test this experimentally, we performed temperature--dependent magnetic susceptibility measurements on a piece of a combinatorial film at several applied fields down to $T=2$ K, as shown in Fig. \ref{fgr:prop}a. The cation composition of this sample, as measured by XRF, is approximately Gd/(W+Gd) $\approx43$\% (Fig. S8). The diamagnetic contribution from the Si substrate has been measured and subtracted, and the sample mass was calculated from the approximate film volume (see Supporting Information for details). The temperature-dependent susceptibility ($\chi$) is consistent with paramagnetism down to $T=2$ K: there is no peak, and no splitting between zero-field-cooled and field-cooled data is observed at low applied field. The inset of Fig. \ref{fgr:prop}a shows magnetization ($M$) as a function of applied magnetic field at $T=2$ and $300$ K. These data are also consistent with paramagnetism; no net moment is observed. Additional temperature- and applied field-dependent data are shown in Fig. S10. 

A Curie-Weiss fit performed from 50 -- 350 K on inverse susceptibility data collected with applied field $\mu_0H=7$ T yields a Weiss temperature of -4.7(1) K and an effective moment ($\mu_{eff}$) of 8.6 $\mu_B$ per Gd (see Fig. S10b; a diamagnetic correction $\chi_0 = 7.436\times10^{-4}$ was used). This $\mu_{eff}$ value is very close to the value expected for one isolated \ce{Gd^{3+}} (8 $\mu_B$), especially considering the many sources of potential error in quantifying the mass of the film. The slightly negative Weiss temperature indicates weak AFM correlations, consistent with previous and current theoretical predictions.\cite{Gui2022} However, the data point to a lack of long-range order even at $T=2$ K. Despite the theoretical predictions of order,\cite{Flores-Livas2019,Gui2022} the paramagnetism is not surprising given that the magnetic behavior of the only other reported nitride perovskites with a magnetic \textit{A} cation---\ce{Ce\textit{M}N_{3-y}} with \textit{M} = W, Mo---is somewhat similar.\cite{Sherbondy2022} \ce{CeWN_{3-y}} is paramagnetic down to $T=2$ K with a large, negative Weiss temperature indicating strong AFM correlations, and \ce{CeMoN_{3-y}} is a frustrated AFM below $T_N\approx8$ K.\cite{Sherbondy2022} However, we note that the $A$ = Ce case is much more complicated than the $A$ = Gd case, as Ce has multiple possible formal oxidation states---as do Mo and W---while Gd does not. The simpler paramagnetism of \ce{GdWN_{3-y}}, in which only the $A$ site is magnetic, compared to \ce{Ce\textit{M}N_{3-y}} suggests that the low-temperature AFM order observed in \ce{CeMoN_{3-y}} may be due to complex coupling between moments on both the \textit{A}-site and \textit{B}-site sublattices. Future strategies to engender long-range magnetic order in nitride perovskites may include increasing distortion in the perovskite lattice, as the \ce{Gd^{3+}} cations are quite far from each other (nearly 4 \AA) in the structural models tested here, or designing materials with magnetic cations on both the \textit{A}- and \textit{B}-site sublattices.

Optical properties were probed with UV-vis transmission spectroscopy, and the extracted absorption coefficients ($\alpha$) are shown in Fig. \ref{fgr:prop}b. The data across a wide composition gradient show a significant change in behavior: the most Gd-rich samples have a  absorption onset of roughly $\sim$3 eV. Fringes are visible in the data for Gd-rich regions, indicating uncorrected reflections. We note that the measured film was approximately 700 -- 1000 nm thick, with a thickness gradient corresponding to the composition gradient. As Gd/(W+Gd) decreases, the absorption onset also decreases. Close to the point where the perovskite phase crystallizes, the fringes disappear and the behavior becomes highly absorbing across a wide energy range. GIWAXS data and an optical image of the sample on which UV-vis was measured are shown in Fig. S7. The film is transparent yellow-brown in the Gd-rich region, above approximately Gd/(W+Gd) $\approx60$\%. Below this, the film is dark, consistent with the higher absorption and possible impurities of metallic W or \ce{WN_x}. Additionally, spectroscopic ellipsometry data were collected on an as-deposited and an annealed, crystalline perovskite film (Fig. S11). The as-deposited data were easily modeled as an amorphous semiconductor, while the annealed, crystalline samples were much more difficult to model due to high optical absorption. The absorption increases upon annealing, particularly below $\sim3$ eV. The trends are consistent with those extracted from the UV-vis data at similar Gd/(W+Gd) values.

We note that the reported nitride perovskites are dominated by W as the \textit{B}-site cation; only one example with \textit{B} = Mo, which is directly above W in the periodic table and should presumably follow similar trends, has been reported. Indeed, our attempts to synthesize \ce{GdMoN3} and \ce{LaMoN3} in a similar manner to \ce{LaWN3},\cite{Talley2021,Smaha2023} \ce{CeWN3}, \ce{CeMoN3},\cite{Sherbondy2022}, and now \ce{GdWN3} have been unsuccessful to date. We attempted RF co-sputtering with a range of experimental parameters, including varying the substrate temperature, deposition rate, activation of nitrogen via a nitrogen plasma cracker, biasing the substrate, and annealing post-situ, and did not observe the crystallization of a perovskite phase. This may be consistent with computational work predicting that the most stable structure of \ce{LaMoN3} is not a perovskite and predicting that pressure of approximately 1.5 GPa is necessary to stabilize a perovskite phase.\cite{Gui2020} However, similar calculations predicted analogous behavior for \ce{GdWN3},\cite{Gui2022} which we show experimentally here did form a perovskite. The films resulting from our experimental attempts at \ce{GdMoN3} and \ce{LaMoN3} were generally highly air-sensitive and were often complex mixes of binary nitrides, oxides, and oxynitrides. \ce{CeMoN_{3-y}} is the only Mo-containing nitride perovskite that has been successfully synthesized;\cite{Sherbondy2022} we hypothesize that it may be stabilized by the possibility of mixed \ce{Ce^{3+}}/\ce{Ce^{4+}} oxidation states compensated by \ce{Mo^{5+}}/\ce{Mo^{6+}}. 

In conclusion, we present the synthesis and characterization of the first Gd-containing nitride perovskite, \ce{GdWN_{3-y}}. Combinatorial RF co-sputtering with activated nitrogen yielded nanocrystalline films with a composition gradient. AES depth profiling and STEM-EDS elemental mapping reveal low oxygen contents in the bulk of the film, consistent with a primarily nitride compound. In contrast with previous ground-state structure calculations, an ex-situ rapid thermal annealing step in flowing \ce{N2} crystallized a perovskite phase close to the \ce{GdWN_{3-y}} composition. LeBail fits of synchrotron GIWAXS data are most consistent with the calculated ground-state \textit{Pnma} perovskite structure. Density functional theory calculations reveal a perovskite ground state of \ce{GdWN3} in orthothombic space group \textit{Pnma} with AFM magnetic ordering of Gd$^{3+}$, in contrast to previous computational studies predicting a non-perovskite structure.  Magnetic susceptibility measurements reveal paramagnetic behavior down to $T=2$ K with an effective moment of 8.4 $\mu_B$, indicating isolated \ce{Gd^{3+}} moments. The negative Weiss temperature suggests AFM correlations, consistent with calculations of an AFM ground state. Optical properties were studied across the composition gradient, finding an increase in the absorption onset in Gd-rich samples and a decrease in W-rich samples. Future work will be required to further explore and optimize this new material. The high-throughput discovery and characterization of \ce{GdWN3}, the first Gd-containing nitride perovskite, opens a pathway for further understanding, designing, and controlling the properties of this emerging class of materials.

\begin{suppinfo} 
Experimental methods and additional AES, TEM, SEM, XRD, calculations, magnetism, and spectroscopic ellipsometry data (PDF)
\end{suppinfo}

\begin{acknowledgement}
This work was authored by the National Renewable Energy Laboratory (NREL), operated by Alliance for Sustainable Energy, LLC, for the U.S. Department of Energy (DOE) under Contract No. DE-AC36-08GO28308. This work was supported by the Laboratory Directed Research and Development (LDRD) Program at NREL (R.W. Smaha, calculations). Funding was provided by the U.S. Department of Energy, Office of Science, Basic Energy Sciences, Division of Materials Science: through Materials Chemistry program, as a part of the Early Career Award ``Kinetic Synthesis of Metastable Nitrides'' (development and operation of synthesis and characterization equipment); through the Funding Opportunity Announcement (FOA) DE-FOA-0002676: Chemical and Materials Sciences to Advance Clean-Energy Technologies and Transform Manufacturing (magnetic measurements), and through the Laboratory Program Announcement LAB 21-2491: Microelectronics Co-Design Research (microscopy). This work used computational resources sponsored by the Department of Energy’s Office of Energy Efficiency and Renewable Energy, located at NREL. Use of the Stanford Synchrotron Radiation Lightsource, SLAC National Accelerator Laboratory, is supported by the U.S. Department of Energy, Office of Science, Office of Basic Energy Sciences under Contract No. DE-AC02-76SF00515. The authors wish to thank the Analytical Resources Core (RRID: SCR\_021758) at Colorado State University for instrument access, training and assistance with sample analysis. The authors thank N. Strange for support with GIWAXS measurements and P. Walker for assistance with the FIB liftout. The views expressed in the article do not necessarily represent the views of the DOE or the U.S. Government. 
\end{acknowledgement}

\textbf{Author Contributions}:

\textbf{Rebecca W. Smaha}: Conceptualization, Investigation, Formal Analysis, Data Curation, Writing - Original Draft, Project Administration, Funding Acquisition.
\textbf{John S. Mangum}: Investigation, Formal Analysis, Data Curation, Writing - Original Draft.
\textbf{Neha Yadav}: Investigation, Formal Analysis, Writing - Review \& Editing.
\textbf{Christopher L. Rom}: Investigation, Writing - Review \& Editing.
\textbf{Brian M. Wieliczka}: Investigation. 
\textbf{Baptiste Julien}: Investigation.
\textbf{Andrew Treglia}: Investigation.
\textbf{Craig L. Perkins}: Investigation.
\textbf{Prashun Gorai}: Formal Analysis, Data Curation, Writing - Review \& Editing.
\textbf{Sage R. Bauers}: Methodology, Writing - Review \& Editing, Supervision.
\textbf{Andriy Zakutayev}: Conceptualization, Methodology, Resources, Writing - Review \& Editing, Supervision, Funding Acquisition.

\providecommand{\latin}[1]{#1}
\makeatletter
\providecommand{\doi}
  {\begingroup\let\do\@makeother\dospecials
  \catcode`\{=1 \catcode`\}=2 \doi@aux}
\providecommand{\doi@aux}[1]{\endgroup\texttt{#1}}
\makeatother
\providecommand*\mcitethebibliography{\thebibliography}
\csname @ifundefined\endcsname{endmcitethebibliography}  {\let\endmcitethebibliography\endthebibliography}{}

\end{document}